\documentclass[a4paper,12pt]{article}

\title{\bf Possible evolution of dim
radio quiet neutron star 1E 1207.4-5209 based on a B-decay model}
\author{A\c{s}k\i n Ankay
\thanks{askin.ankay@boun.edu.tr},
Arzu Mert Ankay
\thanks{arzu.mert@boun.edu.tr},
Enise Nihal Ercan
\thanks{ercan@boun.edu.tr}
\\ \\
{Bo\u{g}azi\c{c}i University} \\
{Department of Physics} \\
{34342-Bebek, \.Istanbul, Turkey}}

\begin{document}

\maketitle

\begin{abstract}
Dim radio-quiet neutron star (DRQNS) 1E 1207.4-5209 is one of the
most heavily examined isolated neutron stars. Wide absorption
lines were observed in its spectrum obtained by both XMM-Newton
and Chandra X-ray satellites. These absorption lines can be
interpreted as a principal frequency centered at 0.7 keV and its
harmonics at 1.4, 2.1 and possibly 2.8 keV. The principal line can
be formed by resonant proton cyclotron scattering leading to a
magnetic field which is two orders of magnitude larger than the
perpendicular component of the surface dipole magnetic field (B)
found from the rotation period (P) and the time rate of change in
the rotation period (\.{P}) of 1E 1207.4-5209. Besides, age of the
supernova remnant (SNR) G296.5+10.0 which is physically connected
to 1E 1207.4-5209 is two orders of magnitude smaller than the
characteristic age ($\tau$=P/2\.{P}) of the neutron star. These
huge differences between the magnetic field values and the ages
can be explained based on a B-decay model. If the decay is assumed
to be exponential, the characteristic decay time turns out to be
several thousand years which is three orders of magnitude smaller
than the characteristic decay time of radio pulsars represented in
an earlier work. The lack of detection of radio emission from
DRQNSs and the lack of point sources and pulsar wind nebulae in
most of the observed SNRs can also be partly explained by such a
very rapid exponential decay. The large difference between the
characteristic decay times of DRQNSs and radio pulsars must be
related to the differences in the magnetic fields, equation of
states and masses of these isolated neutron stars. 

\end{abstract}

\section{Introduction}

Today, several different types of isolated neutron stars which
have comparably very different physical properties are known.
These are radio pulsars, dim radio-quiet neutron stars (DRQNSs),
dim thermal neutron stars (DTNSs), anomalous X-ray pulsars (AXPs)
and soft gamma-ray repeaters (SGRs). About 1700 radio pulsars have
been observed up to date $^1$, some of which have also been
detected in other bands of the electromagnetic spectrum. Radio
pulsar is relatively the most well known and understood object
both theoretically and observationally among all the classes of
isolated neutron stars.

DRQNSs, DTNSs and AXPs-SGRs are new types of isolated neutron
stars (point X-ray sources). All AXPs-SGRs, which should be
considered as a single class, and some DRQNSs and DTNSs, which are
cooling neutron stars with short (P$\sim$few 100 milliseconds) and
long (P$\sim$few seconds) rotation periods respectively, show
themselves as X-ray pulsars. AXPs and SGRs (which have also long
rotation periods similar to DTNSs and which experience X-ray and
$\gamma$-ray bursts) may be in early phases of at least some of
the DTNSs based on field decay models (see Fig.1). Recently, Ankay
et al. $^2$ have shown the possibility that radio-quiet strong
X-ray pulsar J1846-0258 may be in a phase preceding AXP-SGR and/or
DTNS phases based on a detailed comparison between the
observational characteristics of these sources.

None of the DRQNSs, DTNSs or AXPs/SGRs has been detected at radio
frequencies with the exception AXP XTE J1810-197 from the
direction of which radio emission at 1.4 GHz has been detected
$^3$, but this radio emission may also be produced by a possible
pulsar wind nebula (PWN) around this AXP.

There are only a few observationally known DRQNSs. All of them are
physically connected to Galactic supernova remnants (SNRs) which
have ages less than about 20 kyr $^{4,5}$.

There are two basic observable quantities of isolated pulsars: the
rotation period P and the time rate of change of the rotation
period \.{P}. Time rate of change of the rotational kinetic energy
of a rigid body is
\begin{equation}
\dot{E} = I\Omega\dot{\Omega} = \frac{4\pi^2I\dot{P}}{P^3}
\end{equation}
where I is the moment of inertia and $\Omega$=2$\pi$/P is the
rotational velocity. The component of the dipole magnetic field of
pulsar which is perpendicular to the rotation axis also depends on
P and \.{P}:
\begin{equation}
B = (\frac{3c^3IP\dot{P}}{8\pi^2R^6})^{1/2}
\end{equation}
where c is the speed of light in free space and R is the radius of
pulsar.

If we assume that the time rate of change of the rotational
velocity of pulsar can be expressed as a power law:
\begin{equation}
\dot{\Omega} = -k\Omega^n
\end{equation}
where k is a proportionality constant and the power n is called
the 'braking index', then the real age of pulsar can be
represented in terms of P and \.{P} as:
\begin{equation}
t = \frac{P}{(n-1)\dot{P}}[1-(\frac{P_0}{P})^{n-1}]
\end{equation}
where P$_0$ is the initial rotation period of pulsar. If P$_0$ is
much less than P:
\begin{equation}
t\cong\frac{P}{(n-1)\dot{P}}
\end{equation}

Magneto-dipole radiative power of pulsars is
\begin{equation}
L = \frac{B_p^2R^6\Omega^4}{6c^3}Sin^2\alpha
\end{equation}
where B$_p$ is the strength of the dipole magnetic field at the
magnetic pole and $\alpha$ is the angle between the rotation axis
and the magnetic axis $^6$. If \.{E} (eqn.(1)) is equal to L
(eqn.(6)) (i.e. if the net torque on pulsar is equal to the
magneto-dipole radiation torque), then the braking index turns out
to be n=3 (assuming also that eqn.(3) can be applied to express
the rotational evolution of pulsar). In such a case, the
characteristic age ($\tau$) of pulsar is defined using the n=3
condition in eqn.(5):
\begin{equation}
\tau \cong \frac{P}{2\dot{P}}
\end{equation}
So, when n=3 and P$_0$$\ll$P, $\tau$ is approximately equal to the
real age of pulsar. When n is greater (less) than 3, $\tau$ is
greater (less) than the real age. So, if there are extra torques
on pulsar which spin it down in addition to the effect of the
magneto-dipole radiation torque, the pulsar will evolve with
larger \.{P} values compared to the case of pure magneto-dipole
radiation torque. On the other hand, if there is B-decay, the
\.{P} values the pulsar has throughout the evolution will be
smaller. Note that the braking index n should be adapted as the
average braking index ($\bar{n}$) of pulsar when considering long
time intervals in pulsar's lifetime, because the value of
'instantaneous' (i.e. considering very short time intervals
compared to the lifetime of pulsar) braking index may change in
time in a complicated way in general.

Although, none of the DRQNSs has been detected at radio
frequencies, this does not necessarily mean that they have no
radio emission. There are basically two selection effects (other
than the background radiation which is effective only on the
sources located in the Galactic central directions, see Ankay et
al. $^7$) which can prevent detection of the radio emission: the
beaming fraction and the luminosity function. Based on radio
pulsar observations, there exists a strong evidence that beaming
fraction decreases in time possibly because of the angle between
the rotation axis and the magnetic axis ($\alpha$) decreasing
during the evolution and the large number of the observed radio
pulsars with possibly small values of $\alpha$ supports this
interpretation $^8$. Indeed, the existence of B-decay (where B is
the component of the dipole magnetic field perpendicular to the
rotation axis) for a large sample of radio pulsars was shown by
Guseinov et al. $^9$ by comparing the characteristic ages of radio
pulsars with their kinematic ages (i.e. their distances from the
Galactic plane). Guseinov et al. $^9$ assumed an exponential decay
with a characteristic decay time $\tau_d$=3$\times$10$^6$ yr. The
cause of B-decay in the case of radio pulsars can be a temporal
decrease in $\alpha$, but the possibility of a decay in the dipole
magnetic field itself can not be totally excluded (see e.g Geppert
\& Rheinhardt $^{10}$ on the possibility of magnetic field decay
in neutron stars).

The other selection effect is related to radio luminosity versus
number distribution of radio pulsars (luminosity function). Based
on the observational data, most of the radio pulsars must have low
radio luminosity at birth and the radio luminosity, which is only
a small fraction of the magneto-dipole radiation produced by the
pulsar, does not change significantly in time $^{11}$.

As mentioned above, all the known DRQNSs are connected to Galactic
SNRs with ages $\le$2$\times$10$^4$ yr. Most of the Galactic SNRs
are shell-type $^{12-15}$ and many of these SNRs lack detected
point sources or PWNe in them. Direct detection of neutron stars
in many of the SNRs may not be possible because of the selection
effects and the large distances. On the other hand, PWN is not
seen around pulsars which have rate of rotational energy loss
\.{E}$<$10$^{35}$ erg/s $^{5,16,17}$. The only known DRQNS with
measured P and \.{P} values is 1E 1207.4-5209 and it has
\.{E}=2$\times$10$^{34}$ erg/s. This explains why there is no PWN
around 1E 1207.4-5209.

For 1E 1207.4-5209, there are 2 problems based on its position on
the P-\.{P} diagram (Fig.1): first of all, its present B value is
much less than the magnetic field measured from its 0.7 keV proton
cyclotron line detected by both {\it Chandra} and {\it
XMM-Newton}. Secondly, the $\tau$ value of this DRQNS is much
larger than the age of the SNR which it is physically connected
to. Below, we will present a B-decay model in order to clarify
these problems and to explain the possible evolution of 1E
1207.4-5209.

In Section 2, the observational data on 1E 1207.4-5209 are
represented. In Section 3, an exponential B-decay model is
introduced as a possible evolutionary scenario for this DRQNS. In
the last section, discussions and conclusions are represented.

\section{The observational data on 1E 1207.4-5209}

DRQNS 1E 1207.4-5209 is the only known isolated neutron star which
have X-ray absorption lines clearly observed in its spectrum. The
possible proton cyclotron line (0.7 keV) of 1E 1207.4-5209 and its
harmonics at 1.4 and 2.1 keV (and possibly at 2.8 keV) detected by
{\it XMM-Newton} (260 ks observation $^{18,19}$) corresponds to a
magnetic field $\sim$1.6$\times$10$^{14}$ G $^{18,19}$. Sanwal et
al. $^{20}$ also pointed out the possibility that the 0.7 and 1.4
keV lines ({\it Chandra} observations $\sim$60 ks) could be formed
by transitions of singly ionized Helium in a magnetic field
$\sim$(1.4-1.7)$\times$10$^{14}$ G. We will adapt
B$_{cyclotron}$=1.6$\times$10$^{14}$ G as the present surface
dipole magnetic field of 1E 1207.4-5209. Note that, if the
cyclotron line is interpreted in terms of electrons, the
corresponding magnetic field value ($\sim$8$\times$10$^{10}$ G
$^{18,19}$) is much smaller than its present B value.

Assuming conventional values of mass and radius for 1E 1207.4-5209
and using its measured P=424 ms and
\.{P}$\cong$1.4$\times$10$^{-14}$ ss$^{-1}$ values, one finds
B$\cong$2.5$\times$10$^{12}$ G. This leads to a very small angle
between the magnetic axis and the rotation axis:
$\alpha\cong$0$^o$.9. On the other hand, SNR G296.5+10.0, which is
physically connected to 1E 1207.4-5209, has age t$_{SNR}\cong$ 10
kyr $^{4,5,15}$ and the characteristic age of 1E 1207-5209
$\tau\cong$480 kyr. The difference between t$_{SNR}$ and $\tau$
leads to a very large average braking index unless the initial
rotation period (P$_0$) of the pulsar is comparable to its present
P value. If 1E 1207.4-5209 has evolved with n=3 since it was born,
it should have P$_0\cong$420 ms (see eqn.4). A neutron star born
with such a large rotation period is a significant problem for
theories on core-collapse supernovae and formation of pulsars (see
Ardeljan et al. $^{21}$, Moiseenko et al. $^{22}$ and Heger et al.
$^{23,24}$ on the necessity of rapid rotation for newborn neutron
stars), though this possibility may not be totally excluded.

\section{B-decay model for 1E 1207.4-5209}

We can explain the huge difference between t$_{SNR}$ and $\tau$
and the large discrepancy between B$_{cyclotron}$ and B by a
simple B-decay model. We will assume an exponential decay similar
to Guseinov et al. $^9$ who showed the existence of B-decay for a
large sample of radio pulsars by comparing the kinematic ages with
the characteristic ages. Assuming conventional values of mass and
radius for the pulsar and using eqn.(2), the exponential B-decay
can be written as:
\begin{equation}
P \dot{P} = \frac{B_0^2}{10^{39}} e^{-2t/\tau_d}
\end{equation}
for t$\gg$$\tau_d$. Here, P and \.{P} are the present values of
the pulsar, B$_0$ is the initial value of the magnetic field
component perpendicular to the rotation axis, $\tau_d$ is the
characteristic decay time, and t is the real age of pulsar. If the
initial angle $\alpha_0$=90$^o$ (i.e. if B$_0$=B$_{cyclotron}$
assuming no field decay), then $\tau_d\cong$2.4$\times$10$^3$ yr
for t=t$_{SNR}\cong$10$^4$ yr.

We can also put an upper limit on $\tau_d$: if the pulsar had
evolved with n=3, it should have had
B$_0\cong$1.7$\times$10$^{13}$ G (assuming that the pulsar's age
is 10$^4$ yr and P$_0\ll$P, see Fig.1). Using
B$_0$=2$\times$10$^{13}$ G in eqn.(8) gives
$\tau_d\cong$4.8$\times$10$^3$ yr. So,
$\tau_d\cong$(2.4-4.8)$\times$10$^3$ yr corresponding to
B$_0\cong$(16-2)$\times$10$^{13}$ G. In this analysis, we assume
that a possible evolution with $\bar{n}<$3 when the pulsar is very
young compared to its present age will have a negligible influence
on its evolution on the P-\.{P} diagram.

In the B-decay model introduced above, the characteristic age can
be written as a function of $\tau_d$ and t as:
\begin{equation}
\tau = \frac{\tau_d}{2} e^{2t/\tau_d}
\end{equation}
with the condition t$\gg$$\tau_d$. Using eqns. (7),(8) and (9), we
have found the positions on the P-\.{P} diagram (shown by 'light
squares' in Fig.1) corresponding to different evolutionary tracks
based on the B-decay model. In Fig.1, the points ('light squares')
a (t=5 kyr), b (t=10 kyr) and c (t=15 kyr) are on the evolutionary
track with B$_0$=2$\times$10$^{13}$ G and $\tau_d$=2.5 kyr. The
points d (t=10 kyr), e (t=15 kyr) and f (t=20 kyr) correspond to
the evolutionary track with B$_0$=2$\times$10$^{13}$ G and
$\tau_d$=5 kyr. The points g (t=5 kyr), h (t=10 kyr), i (t=15 kyr)
and j (t=10 kyr), k (t=15 kyr), l (t=20 kyr) are on the
evolutionary tracks with B$_0$=4$\times$10$^{13}$ G, $\tau_d$=2.5
kyr and B$_0$=4$\times$10$^{13}$ G, $\tau_d$=5 kyr, respectively.

The present position of 1E 1207.4-5209 on the P-\.{P} diagram is
in between the last two tracks (B$_0$=4$\times$10$^{13}$ G,
$\tau_d$=2.5 kyr and B$_0$=4$\times$10$^{13}$ G, $\tau_d$=5 kyr,
see Fig.1). We can conclude that B$_0\cong$4$\times$10$^{13}$ G
and $\tau_d\cong$3.75 kyr for this DRQNS in the B-decay model. The
corresponding initial value of the angle between the rotation axis
and the magnetic field axis is $\alpha_0\cong$14$^o$.5.

\section{Discussions and Conclusions}

If the present surface dipole magnetic field of 1E 1207.4-5209 is
about 1.6$\times$10$^{14}$ G based on resonant proton cyclotron
scattering measurements and since B$_{1207}$=2.5$\times$10$^{12}$
G, there may be a decay in the angle $\alpha$ between the rotation
and the magnetic axes (but not a decay in the magnetic field
itself) considering the huge difference between the characteristic
age of 1E 1207.4-5209 and the age of its SNR. This can be
explained by a simple exponential B-decay model. Characteristic
time of the exponential decay must be about 2.5-5 kyr depending on
the initial value of B which must be in the interval
B$_0$=(16-2)$\times$10$^{13}$ G. Such a rapid decrease in $\alpha$
leads to a short lifetime (possibly about (1-2)$\times$10$^4$ yr
taking into consideration the ages of the SNRs physically
connected to DRQNSs) for this type of neutron star as an X-ray
pulsar and this may also be the reason for the lack of detected
radio emission from such neutron stars as the radio beam width can
be narrower compared to the X-ray beam width. After about
(1-2)$\times$10$^4$ yr pass since the birth of such a pulsar, it
can be detected as a cooling neutron star (but not as an X-ray or
radio pulsar) if it is located close enough to the Sun. This
result can be used to explain the lack of detected point sources
in at least some of the Galactic SNRs (especially the distant
ones) as the lifetime of an SNR can be as long as 10$^5$ yr.

The huge difference between the characteristic age of 1207.4-5209
and the age of its SNR may in principle be explained by assuming
that this DRQNS was born with an initial rotation period very
close to its present rotation period. But such a long initial
rotation period (P$_0$$\cong$420 ms) contradicts the estimations
based on existing theories on formation of neutron stars by
core-collapse supernovae (which predict one order of magnitude
smaller P$_0$ values).

On the other hand, the huge difference between the surface dipole
magnetic field (about 1.6$\times$10$^{14}$ G obtained from the
cyclotron line measurement) of 1E 1207.4-5209 and its B value
(about 2.5$\times$10$^{12}$ G) can not be explained by assuming a
constant and very small initial angle ($\alpha_0$$\cong$0$^o$.9)
between the rotation and magnetic axes, if the initial rotation
period is much less than the present rotation period (i.e. if the
average braking index is much greater than 3, see eqn.4 and
eqn.5).

There may be basically two different magnetic fields in neutron
stars: the magnetic field formed by superconducting entrainment
currents which is parallel to the rotation axis and the 'relic'
magnetic field (which may be homogeneous inside and in the form of
a dipole field on the surface of the neutron star) which is formed
when the progenitor star collapses leading to supernova explosion
$^{25,26}$. The relic magnetic field axis can have an arbitrary
angle ($\alpha$) with respect to the rotation axis and the
resultant magnetic field will be a superposition of the two
magnetic fields.

As the axis of the relic magnetic field approaches the rotation
axis, the magnitude of the resultant surface magnetic field (i.e.
the total magnetic field on the surface which is the superposition
of the relic field and the field formed by the superconducting
entrainment currents) increases and its component which is
perpendicular to the rotation axis decreases in time. Since 1E
1207.4-5209 has most probably very high B$_{cyclotron}$ and
comparably very low B (perpendicular component of the resultant
surface magnetic field) and as this neutron star is observed as a
pulsar, B$_{cyclotron}$ must be comparable to the resultant
surface magnetic field which is increasing in time (or to the
relic magnetic field as the magnetic field formed by the
entrainment currents must be on the order of 10$^{12}$ G on
average $^{25,26}$). On the other hand, the actual low B value of
1E 1207.4-5209 as compared to the case of n=3 (see Fig.1) can be
explained by the exponential decay as this perpendicular component
decreases in time based on the discrepancy between the age values.

The increase in the magnitude of the resultant surface magnetic
field of 1E 1207.4-5209 must be very small if the relic field is
actually much higher than the entrainment field, so that the
B$_{cyclotron}$ and hence the magnitude of the present resultant
surface magnetic field must be comparable to the magnitude of the
initial resultant surface magnetic field. On the other hand, the
decrease in the perpendicular component of the resultant surface
magnetic field must be very large (about one order of magnitude)
in a very short time interval ($\sim10^4$ yr) compared to radio
pulsar lifetimes. In the case of radio pulsars, the increase in
the magnitude of the resultant surface magnetic field must be
relatively large (up to a factor of about 1.5) based on the known
B values of radio pulsars and the prediction on the average
strength of the entrainment field, and the decrease in the
perpendicular component of the field must exist on a much longer
timescale. Comparing the magnetic field values of 1E 1207.4-5209
with the B values of most of the radio pulsars together with the
age discrepancy for 1E 1207.4-5209 -- SNR G296.5+10.0 pair (which
also exists for several other pulsar -- SNR pair but without any
clearly detected cyclotron line) may be an evidence for the
existence of two different magnetic fields for neutron stars; the
relic field and the entrainment field. The interaction between the
superconducting entrainment currents and the relic field may be
the reason for the decrease in $\alpha$. Such a possible
interaction between the two fields throughout the evolution of
neutron stars with different initial intrinsic physical conditions
has yet to be examined.

The rate of temporal decrease in $\alpha$ (and hence the evolution
of the perpendicular component of the total surface dipole
magnetic field) may depend on both the relic and hence the
resultant magnetic field, the mass and the equation of state of
the neutron star (it may also depend on the initial value of
$\alpha$). The changes in \.{P} observed for 1E 1207.4-5209 may
also be explained by oscillations in the magnetic dipole axis as
it approaches the rotation axis. Such changes in \.{P} must also
exist in other 1207-like isolated radio-quiet X-ray pulsars.

As a comparison please note that the magnetic field values of
accreting X-ray pulsars in X-ray binaries obtained from cyclotron
line measurements are on the order of 10$^{12}$ G $^{27}$, that is
comparable to the predicted surface magnetic fields due to the
entrainment currents $^{25,26}$, and the B values of recycled
millisecond pulsars are several orders of magnitude smaller (see
Bisnovatyi-Kogan \& Komberg $^{28,29}$ who gave the first reliable
explanation for the magnetic field decay in X-ray binaries by
accretion and the formation of recycled millisecond pulsars). The
surface magnetic field in the case of X-ray binaries most probably
decreases because of the plasma falling upon the surface of the
neutron star during the accretion process $^{28-32}$. A decrease
in the resultant surface field because of accretion and maybe also
a decrease in the value of $\alpha$ (but on a much longer
timescale compared to the 1E 1207.4-5209 case) may explain the
measured conventional 10$^{12}$ G magnetic field values of X-ray
pulsars in binary systems and the very low perpendicular
components of the resultant surface field of recycled millisecond
pulsars.

As a last remark we would like to note that the birth rate of
1207-like X-ray pulsars must be about 20-30\% of the total
supernova rate (excluding type-Ia supernovae as this type of
explosion most probably does not lead to formation of a neutron
star) taking into consideration that the lifetime of such sources
as X-ray pulsars must be about 2$\times$10$^4$ yr because of rapid
B-decay (see Guseinov et al. $^5$ on the birth rates of different
types of isolated neutron stars including dim radio quiet neutron
stars some of which seem to be 1207-like objects based on the
existing observational data).

{\bf Acknowledgments} We would like to thank the anonymous referee
for his/her constructive and helpful contributions. This work is
supported by Bo\u{g}azi\c{c}i University.

\section{References}
1. ATNF -- Australia Telescope National Facility Pulsar Catalogue
(2006), \\
http://www.atnf.csiro.au/research/pulsar/psrcat/. \\
2. A. Ankay, S. Sahin, G. Karanfil and E. Yazgan, Int. J. Mod.
Phys. D {\bf 14}, 1075 (2005). \\
3. J. P. Halpern, E. V. Gotthelf, R. H. Becker, D. J. Helfand and
R. L. White, Astrophys. J. {\bf 632}, L29 (2005). \\
4. O. H. Guseinov, A. Ankay and S. O. Tagieva, Astrophys. Space
Sci. {\bf 289}, 23 (2004). \\
5. O. H. Guseinov, A. Ankay and S. O. Tagieva, Int. J. Mod. Phys D
{\bf 14}, 643 (2005). \\
6. V. M. Lipunov, Astrophysics of Neutron Stars, Springer (1992).
\\
7. A. Ankay, O.H. Guseinov and S.O. Tagieva, Astronomical and
Astrophysical Transactions {\bf 23}, 503 (2004). \\
8. T. M. Tauris and R. N. Manchester, Mon. Not. R. Astron. Soc.
{\bf 298}, 625 (1998). \\
9. O. H. Guseinov, A. Ankay and S. O. Tagieva, Int. J. Mod. Phys.
D {\bf 13}, 1805 (2004). \\
10. U. Geppert and M. Rheinhardt, Astronomy and Astrophysics {\bf
392}, 1015 (2002). \\
11. O. H. Guseinov, E. Yazgan, S. O. Tagieva and S. Ozkan, Rev.
Mex. Astron. Astrof. {\bf 39}, 267 (2003). \\
12. D. A. Green, A Catalogue of Galactic Supernova Remnants
(2006 April version), http://www.mrao.cam.ac.uk/surveys/snrs/. \\
13. O. H. Guseinov, A. Ankay and S. O. Tagieva, Serb. Astron. J.
{\bf 167}, 93 (2003). \\
14. O. H. Guseinov, A. Ankay and S. O. Tagieva, Serb. Astron. J.
{\bf 168}, 55 (2004). \\
15. O. H. Guseinov, A. Ankay and S. O. Tagieva, Serb. Astron. J.
{\bf 169}, 65 (2004). \\
16. O. H. Guseinov, A. Ankay, S. O. Tagieva and M. O. Taskin, Int.
J. Mod. Phys. D {\bf 13}, 197) (2004). \\
17. O. H. Guseinov, A. Ankay and S. O. Tagieva, astro-ph/0401114 (2004). \\
18. A. De Luca, P. Caraveo, S. Mereghetti, M. Moroni, G. F.
Bignami and R. Mignani, Young Neutron Stars and Their
Environments, IAU Symposium no.218, edited by Fernando Camilo and
Bryan M. Gaensler, p.273 (2004). \\
19. A. De Luca, S. Mereghetti, P. Caraveo, M. Moroni, R. Mignani
and G. F. Bignami, Astron. Astrophys. {\bf 418}, 625 (2004). \\
20. D. Sanwal, G. G. Pavlov, V. E. Zavlin and M. A. Teter,
Astrophys. J. {\bf 574}, L61 (2002). \\
21. N. V. Ardeljan, G. S. Bisnovatyi-Kogan and S. G. Moiseenko,
Mon. Not. R. Astron. Soc. {\bf 359}, 333 (2005). \\
22. S. G. Moiseenko, G. S. Bisnovatyi-Kogan, N. V. Ardeljan, Mon.
Not. R. Astron. Soc. {\bf 370}, 501 (2006). \\
23. A. Heger, S. E. Woosley, N. Langer and H. C. Spruit, Stellar
Rotation, proceedings of IAU Symposium No.215, edi. by Andre
Maeder and Philippe Eenens, p.591 (2004). \\
24. A. Heger, S. E. Woosley and H. C. Spruit, Astrophys. J. {\bf
626}, 350 (2005). \\
25. D. M. Sedrakian and D. Blaschke, Astrophysics {\bf 45}, 166
(2002). \\
26. D. M. Sedrakian and K. M. Shahabasyan, "The Magnetic Field of
Pulsars" in Neutron Stars, Supernovae and Supernova Remnants, edi.
by Efe Yazgan, Askin Ankay and Oktay H. Guseinov, Nova Science
Publishers (to be published in 2006). \\
27. W. Coburn, W. A. Heindl, R. E. Rothschild, D. E. Gruber, I.
Kreykenbohm, J. Wilms, P. Kretschmar and R. Staubert, Astrophys.
J. {\bf 580}, 394 (2002). \\
28. G. S. Bisnovatyi-Kogan and B. V. Komberg, Sov. Astron.
{\bf 18}, 217 (1974). \\
29. G. S. Bisnovatyi-Kogan and B. V. Komberg, Sov. Astron.
Lett. {\bf 2}, 130 (1976). \\
30. G. S. Bisnovatyi-Kogan and A. M. Fridman, Astron. Zh. {\bf
46}, 721 (1969). \\
31. P. R. Amnuel and O. H. Guseinov, Astron. Tsirk. No.524 (1969).
\\
32. G. S. Bisnovatyi-Kogan, "Evolutions of Neutron Stars and Their
Magnetic Fields" in Neutron Stars, Supernovae and Supernova
Remnants, edi. by Efe Yazgan, Askin Ankay and Oktay H. Guseinov,
Nova Science Publishers (to be published in 2006). \\

\clearpage
\section*{Figure Caption}
Rotation period (P) versus temporal change of rotation period
(\.{P}) diagram of various types of isolated pulsars. Small dots
represent radio pulsars. Symbols 'cross' and 'star' denote
AXPs-SGRs and DTNSs, respectively (upper limits on \.{P} for 3
DTNSs are shown by arbitrary arrows). The actual position of 1E
1207.4-5209 is shown by a 'dark square' and its position for n=3
case (see text) is displayed by a 'plus' sign. Constant lines of
B, \.{E} and $\tau$ are denoted by B11-B15, E29-E41 and T3-T9,
respectively. Some points on the evolutionary tracks found from
the exponential B-decay model (see text) are displayed as 'light
squares': a,b,c (B$_0$=2$\times$10$^{13}$ G, $\tau_d$=2.5 kyr);
d,e,f (B$_0$=2$\times$10$^{13}$ G, $\tau_d$=5 kyr); g,h,i
(B$_0$=4$\times$10$^{13}$ G, $\tau_d$=2.5 kyr); j,k,l
(B$_0$=4$\times$10$^{13}$ G, $\tau_d$=5 kyr).

\clearpage
\begin{figure}[t]
\vspace{21cm} \includegraphics{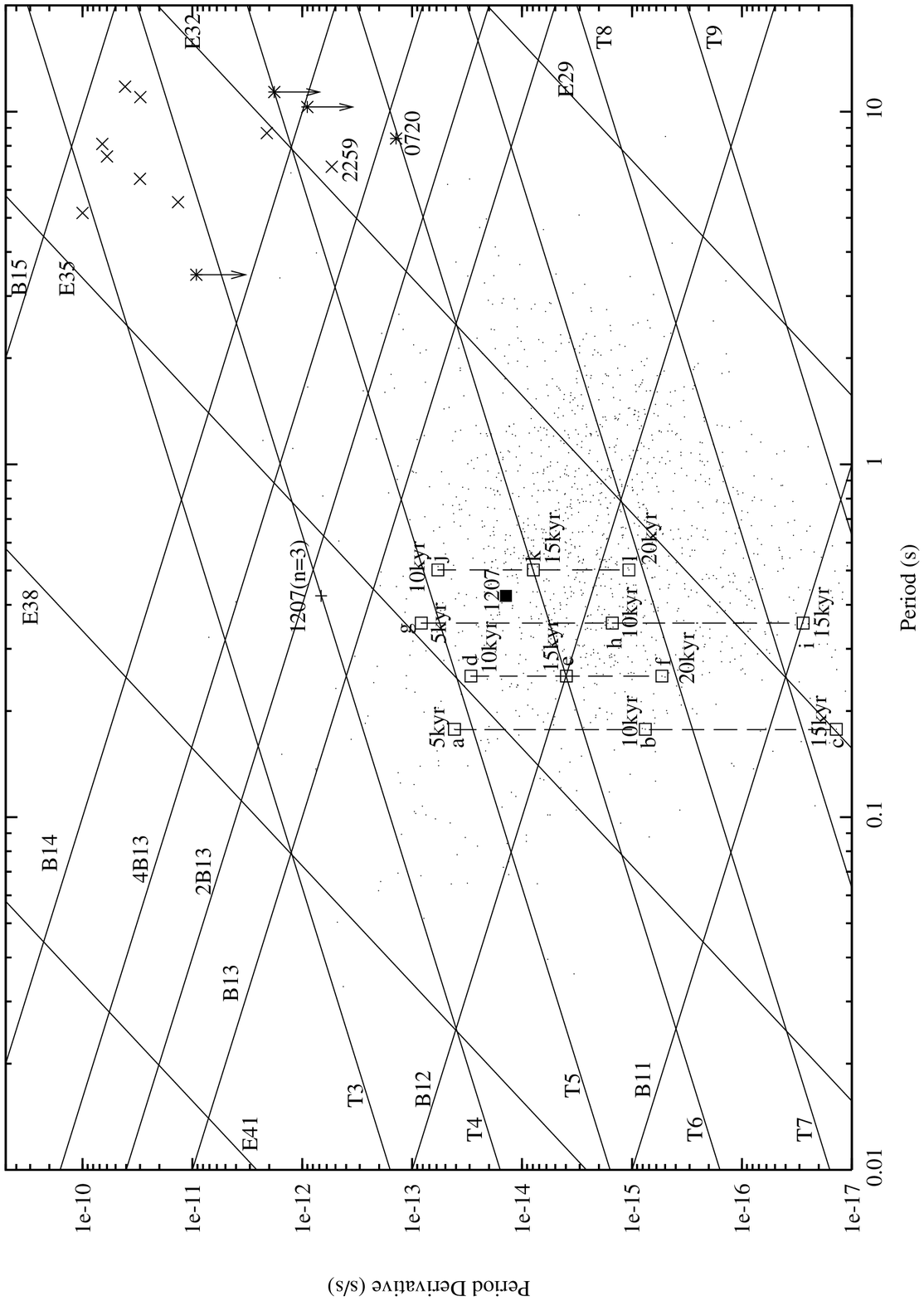}
\end{figure}

\end{document}